\documentclass[aps, two column]{revtex4}
\usepackage{eurosym}
\usepackage{amsfonts}
\usepackage{amsmath}
\usepackage{amssymb,epsf}
\usepackage{color}
\usepackage{graphicx}
\usepackage{epstopdf}
\usepackage{float}
\usepackage{caption}
\usepackage{subfig}
\usepackage{comment}
\usepackage{ulem}

\begin{document}

\title{Black hole solutions in theory of ModMax-dRGT-like massive gravity}
\author{B. Eslam Panah$^{1}$ \footnote{%
email address: eslampanah@umz.ac.ir}}
\affiliation{$^{1}$ Department of Theoretical Physics, Faculty of Basic Sciences, University of Mazandaran, P. O. Box 47416-95447, Babolsar, Iran}

\begin{abstract}
This paper explores the properties of black holes using a new model of
nonlinear electrodynamics called modified Maxwell (ModMax), in conjunction
with nonlinear massive gravity known as dRGT-like massive gravity. We start
by deriving the exact black hole solutions within the framework of
ModMax-dRGT-like massive gravity and analyze how the parameters of both
ModMax and dRGT-like massive gravity influence the characteristics of these
black holes. Additionally, we calculate the thermodynamic quantities for
these black holes in the non-extended phase space and investigate how the
parameters of ModMax and dRGT-like massive gravity affect these quantities.
We confirm that these quantities satisfy the first law of thermodynamics. We also examine local stability by analyzing heat capacity and
assess how ModMax and the massive parameters influence phase transitions and
physical limitation points. Next, we expand our analysis to the extended
phase space, demonstrating that these thermodynamic quantities satisfy both
the first law of thermodynamics and the Smarr relation in this context.
Finally, we examine the isoperimetric ratio of black holes in
ModMax-dRGT-like massive gravity.
\end{abstract}

\maketitle

\section{Introduction }

The black hole is one of the most enigmatic objects
in both theoretical and observational physics. It was first detected in 2016
by the Advanced LIGO/Virgo collaboration through gravitational waves \cite%
{LIGOVIrgoI}. In 2019, the Event Horizon Telescope captured the first image
of a black hole \cite{EHT}. These groundbreaking discoveries impose
constraints on the properties of black holes \cite%
{Cont1,Cont2,Cont3,Cont4,Cont5,Cont6,Cont7}, as well as on modified theories
of gravity. For instance, they suggest a stringent bound on the graviton
mass \cite{Ligo2} and the speed of gravitational waves \cite{gravwave}. This
evidence has motivated an exploration of black holes within the framework of
massive gravity.

Notably, in 2011, de Rham, Gabadadze, and Tolley (dRGT) introduced a
ghost-free nonlinear theory of massive gravity known as dRGT massive gravity 
\cite{dRGTI,dRGTII}. This theory uses a reference metric to construct
massive terms \cite{dRGTI,dRGTII,dRGTIII}, which are integrated into the
action to generate massive gravitons. In 2013, Vegh expanded on the dRGT
massive gravity theory by applying holographic principles and utilizing a
singular reference metric, resulting in dRGT-like massive gravity \cite{Vegh}%
. Recently, this extended version of dRGT massive gravity has attracted
significant interest from various perspectives within the field of physics.

Massive gravity provides an explanation for the acceleration of our
Universe. Recent findings from the advanced LIGO/Virgo collaboration have
established an upper limit on the nonzero mass of gravitons in massive
gravity \cite{Ligo2}. It has been demonstrated that while the mass of
gravity is negligible in typical weak gravity environments, it becomes
significantly more pronounced in strong gravity regimes, such as those
surrounding compact objects and black holes \cite{ZhangZ2018}. This
observation suggests that the mass of the graviton may vary. This theory of
gravity effectively describes the rotation curves of the Milky Way and
spiral galaxies \cite{PanpanichB2018}. It also accounts for the existence of
massive neutron stars \cite{HendiBEP2017}, and predicts the existence of
super-Chandrasekhar white dwarfs \cite{EslamPanahLiu}. Furthermore, quark
stars within this massive gravity framework have been proposed as candidates
for the mass-gap regime \cite{Sedaghat}. Additionally, this theory can
explain current observations related to dark matter \cite%
{Babichev2016a,Babichev2016b}. Instead of specifying an initial ansatz for
the Stueckelberg fields and the reference metric in dRGT massive gravity, it
has been suggested that imposing homogeneity and isotropy on the square root
of the massive tensor leads to dynamic cosmological solutions \cite%
{Heisenberg}. The effects of massive gravitons on topological defects, which
describe magnetic solutions, have been explored in Refs. \cite%
{Magnetic1,Magnetic2,Magnetic3}. Moreover, the existence of remnants for
black holes \cite{RemnantI,RemnantII} is a significant achievement within
the context of dRGT-like massive gravity. It has been shown that topological
black holes in dRGT-like massive gravity can exhibit van der Waals behavior
and critical phenomena, contrasting sharply with their counterparts in
General Relativity \cite{TopologyMS}. Other properties of black holes in
massive gravity have also been investigated in Refs. \cite%
{BH1,BH2,BH3,BH4,BH5,BH6,BH7,BH9,BH10,BH11,BH12,BH13,BH14,BH15,BH16,BH17,BH18,BH19,BH20,BH21,BH22,BH23,BH24,BH25,BH26,BH27}%
.

Nonlinear electrodynamics theories were introduced to address certain issues
in Maxwell's theory of electrodynamics. Two prominent examples are the
Born-Infeld \cite{BI} and Euler-Heisenberg \cite{EH} theories. The
Born-Infeld theory was developed to eliminate the infinite self-energy
associated with the electron's electric field. In contrast, the
Euler-Heisenberg theory provides a complete non-perturbative one-loop
effective action for quantum electrodynamics, accounting for vacuum
polarization effects caused by virtual electrons and positrons. Notably,
Maxwell's theory in four-dimensional spacetime exhibits two remarkable
features: the electromagnetic duality of Maxwell's field equations and the
conformal invariance of the Maxwell action. Therefore, a viable nonlinear
electrodynamics theory, which serves as a nonlinear generalization of
Maxwell's theory, must preserve both of these symmetries. In this context, a
new model of nonlinear electrodynamics, known as the modification of
Maxwell's theory (ModMax theory) \cite{ModMaxI,ModMaxII}, has been
introduced, which retains both symmetries of Maxwell's theory. Several
aspects of this model have been explored in the literature \cite%
{MM1,MM2,MM3,MM4,MM5,MM6,MM7,MM8,MM9,MM10,MM11,MM12}.

\section{Action and Field Equations}

In dRGT massive gravity, considering
the cosmological constant and the ModMax field, the action is given by 
\begin{equation}
I=\frac{1}{\kappa ^{2}}\int d^{4}x\sqrt{-g}[\mathcal{R}+m_{g}^{2}{%
\sum\limits_{i}^{4}c_{i}\mathit{u_{i}(g,f)}}-2\Lambda -4\mathcal{L}],
\label{action}
\end{equation}%
where $\kappa ^{2}=16\pi $. Notably, we set $G=c=1$ in the action (\ref%
{action}), where $c$ is the speed of light and $G$ is the gravitational
constant. Additionally, $g$ refers to the determinant of the metric tensor $%
g_{\mu \nu}$, i.e., $g=\det(g_{\mu \nu })$. Furthermore, $\mathcal{R}$ and $%
m_{g}$ denote the Ricci scalar and the graviton mass, respectively. The constants $c_{i}$ serve as free parameters of the action. Indeed, $c_{i}$'s are arbitrary constants whose values can be determined according to
observational or theoretical considerations. Furthermore, in a
self-consistent massive gravity theory, it may be necessary for all
coefficients to be negative if $m^{2}>0$. However, this paper does not
impose such a restriction, as fluctuations of certain fields with negative
mass squared can remain stable in AdS space, provided that the mass squared
adheres to the corresponding Breitenlohner-Freedman bounds \cite{BH1}. The
quantities ${\mathit{u_{i}}}$ are introduced as symmetric polynomials of the
eigenvalues of the matrix $K_{~~\nu}^{\mu }=\sqrt{g^{\mu \sigma }f_{\sigma
\nu }}$. Here, $g_{\mu \nu }$ is the dynamical metric tensor, and $f_{\mu
\nu }$ is the reference metric. $u_{i}$ are given in the following form 
\begin{equation}
{\mathit{u}}_{i}=\sum_{y=1}^{i}\left( -1\right) ^{y+1}\frac{\left(
i-1\right) !}{\left( i-y\right) !}{\mathit{u}}_{i-y}\left[ K^{y}\right] ,
\label{U}
\end{equation}%
where ${\mathit{u}}_{i-y}=1$, when $i=y$. In addition, $[K]=K_{a}^{a}$ and $%
[K^{n}]=(K^{n})_{a}^{a}$. In the action (\ref{action}), $\mathcal{L}$ is
associated with the Lagrangian of ModMax and is defined in \cite%
{ModMaxI,ModMaxII} 
\begin{equation}
\mathcal{L}=\mathcal{S}\cosh \gamma -\sqrt{\mathcal{S}^{2}+\mathcal{P}^{2}}%
\sinh \gamma ,  \label{ModMaxL}
\end{equation}%
where $\mathcal{S}=\frac{\mathcal{F}}{4}$ and $\mathcal{P}=\frac{\widetilde{%
\mathcal{F}}}{4}$, respectively, are a true scalar, and a pseudoscalar. $%
\gamma $ and $\mathcal{F}=F_{\mu \nu }F^{\mu \nu }$ are, respectively, a
dimensionless parameter (which is known as the ModMax's parameter), and the
Maxwell invariant. It is notable that, $F_{\mu \nu }=\partial _{\mu
}A_{\nu}-\partial _{\nu }A_{\mu }$ is the electromagnetic tensor field
(where $A_{\mu }$ is the gauge potential). Also, $\widetilde{\mathcal{F}}=$ $%
F_{\mu\nu }\widetilde{F}^{\mu \nu }$, and $\widetilde{F}^{\mu \nu }=\frac{1}{%
2}\epsilon _{\mu \nu }^{~~~\rho \lambda }F_{\rho \lambda }$. Notably, the
ModMax's Lagrangian (Eq. (\ref{ModMaxL})) turns to Maxwell theory when $%
\gamma =0$. Whereas we are interested to get the electrically charged black
holes, so we can consider $\mathcal{P}=0$ in the ModMax's Lagrangian (\ref%
{ModMaxL}).

Here, we aim to derive the equations of motion for dRGT-like massive gravity
in the presence of the ModMax field. By varying the action with respect to
the metric tensor $g_{\mu \nu}$ and the gauge potential $A_{\mu}$, we obtain 
\begin{eqnarray}
G_{\mu \nu }+\Lambda g_{\mu \nu }+m_{g}^{2}\chi _{\mu \nu } &=&8\pi \mathrm{T%
}_{\mu \nu },  \label{field eq} \\
\partial _{\mu }\left( \sqrt{-g}\widetilde{E}^{\mu \nu }\right) &=&0,
\label{eq2a}
\end{eqnarray}%
where $G_{\mu \nu }$ is the Einstein tensor and $\chi_{\mu \nu}$ is referred
to as the massive tensor in the following form 
\begin{equation}
\chi _{\mu \nu }=-\sum_{i=1}^{d-2}\frac{c_{i}}{2}\left[ {\mathit{u}}%
_{i}g_{\mu \nu }+\sum_{y=1}^{i}\frac{\left( -1\right) ^{y}i!}{\left(
i-y\right) !}{\mathit{u}}_{i-y}\left[ K_{\mu \nu }^{y}\right] \right] ,
\label{chi}
\end{equation}%
where $d$ is related to the dimensions of spacetime. We work in a $4-$%
dimensional spacetime, so $d=4$. Aditionally, $\mathrm{T}_{\mu \nu }$ is the
stress-energy tensor, which is given by 
\begin{equation}
8\pi \mathrm{T}^{\mu \nu }=2\left( F^{\mu \sigma }F_{~~\sigma }^{\nu
}e^{-\gamma }\right) -2e^{-\gamma }\mathcal{S}g^{\mu \nu },
\end{equation}%
and $\widetilde{E}_{\mu \nu }$ in Eq. (\ref{eq2a}), is defined as 
\begin{equation}
\widetilde{E}_{\mu \nu }=\frac{\partial \mathcal{L}}{\partial F^{\mu \nu }}%
=2\left( \mathcal{L}_{\mathcal{S}}F_{\mu \nu }\right) ,
\end{equation}%
where $\mathcal{L}_{\mathcal{S}}=\frac{\partial \mathcal{L}}{\partial 
\mathcal{S}}$. So, the ModMax field equation (Eq. (\ref{eq2a})), turns to 
\begin{equation}
\partial _{\mu }\left( \sqrt{-g}e^{-\gamma }F^{\mu \nu }\right) =0.
\label{ModMaxf}
\end{equation}

\section{Black Hole Solutions}

We consider a spherically symmetric spacetime with metric signature $(-,+,+,+)$ 
\begin{equation}
ds^{2}=-\psi \left( r\right) dt^{2}+\frac{dr^{2}}{\psi \left( r\right) }%
+r^{2}(d\theta ^{2}+\sin ^{2}\theta d\phi ^{2}),  \label{metric}
\end{equation}%
where $\psi \left( r\right) $ is the metric function. The spatial reference
metric, or spatial fiducial metric, is a suitable option for the black hole
solutions \cite{Vegh,BH1,BH4}. Hence, we suppose 
\begin{equation}
f_{\mu \nu }=diag\left( 0,0,{C}^{2},C^{2}\sin ^{2}\theta \right) ,
\label{reference metric}
\end{equation}%
where $C$ is a positive constant. Using the line element Eq. (\ref{metric})
and the reference metric (\ref{reference metric}), we can determine the
tensor $K_{~~\mu }^{\mu }$ 
\begin{equation}
K_{~~\nu }^{\mu }=diag\left( 0,0,\dfrac{C}{r},\dfrac{C}{r}\right) ,
\label{k tensor}
\end{equation}%
and 
\begin{eqnarray}
(K^{2})_{~~\nu }^{\mu } &=&diag\left( 0,0,\frac{{C}^{2}}{{r}^{2}},\frac{{C}%
^{2}}{{r}^{2}}\right) ,  \notag \\
(K^{3})_{~~\nu }^{\mu } &=&diag\left( 0,0,\frac{{C}^{3}}{{r}^{3}},\frac{{C}%
^{3}}{{r}^{3}}\right) ,  \notag \\
(K^{4})_{~~\upsilon }^{\mu } &=&diag\left( 0,0,\frac{{C}^{4}}{{r}^{4}},\frac{%
{C}^{4}}{{r}^{4}}\right) ,  \label{k power}
\end{eqnarray}%
and also 
\begin{eqnarray}
\lbrack K] &=&\frac{2C}{r},~~~\&~~~[K^{2}]=\dfrac{2C^{2}}{r^{2}},  \notag \\
\lbrack K^{3}] &=&\dfrac{2C^{3}}{r^{3}},~~~\&~~~[K^{4}]=\dfrac{2C^{4}}{r^{4}}%
.  \label{k trace}
\end{eqnarray}

Since we are working in a $4-$dimensional spacetime, the only non-zero
components of $u_{i}$ are $u_{1}$ and $u_{2}$ \cite{BH4}. Therefore, using
Eqs. (\ref{U}), (\ref{k tensor}) and (\ref{k trace}), $u_{i}$ are obtained 
\begin{eqnarray}
{\mathit{u}}_{1} &=&\dfrac{2C}{r},~~~\&~~~{\mathit{u}}_{2}=\dfrac{2C^{2}}{%
r^{2}},  \notag \\
{\mathit{u}}_{i} &=&0,\text{ \ \ when \ \ }i>2.  \label{ui}
\end{eqnarray}

By substituting Eq. (\ref{ui}) into Eq. (\ref{chi}), we can determine the
elements of the massive tensor in the following forms 
\begin{eqnarray}
\chi _{11} &=&\dfrac{C(Cc_{2}+c_{1}r)\psi \left( r\right) }{r^{2}},  \notag
\\
\chi _{22} &=&-\dfrac{C(Cc_{2}+c_{1}r)}{r^{2}\psi \left( r\right) },  \notag
\\
\chi _{33} &=&-\dfrac{Cc_{1}r}{2},  \notag \\
\chi _{44} &=&-\dfrac{Cc_{1}r\sin ^{2}(\theta )}{2}.  \label{chi1}
\end{eqnarray}

To create a radial electric field, we employ the following gauge potential 
\begin{equation}
A_{\mu }=h(r)\delta _{\mu }^{t},  \label{gauge potential}
\end{equation}%
by applying the metric (\ref{metric}) and the gauge potential (\ref{gauge
potential}) along with the ModMax field equation (\ref{eq2a}), we can obtain 
\begin{equation}
2h^{\prime }(r)+rh^{\prime \prime }(r)=0,  \label{heq}
\end{equation}%
where the prime and double prime denote the first and second derivatives
with respect to $r$, respectively. We obtain $h(r)$ from Eq. (\ref{heq}),
which leads to $h(r)=\frac{-q}{r}$. $q$ is an integration constant
associated with the electric charge. The electric field is derived from the
electromagnetic field tensor, which is $E(r)=\frac{q}{r^{2}}e^{-\gamma }$.

Considering the metric (\ref{metric}) and the equations of motion (\ref%
{field eq}), we can derive the following field equations 
\begin{eqnarray}
Eq_{rr} &=&Eq_{tt}=r\psi ^{\prime }(r)+\psi \left( r\right) -1+\Lambda r^{2}
\notag \\
&&-m_{g}^{2}C(c_{2}C+c_{1}r)+\frac{q^{2}e^{-\gamma }}{r^{2}},  \label{GR1} \\
Eq_{\theta \theta } &=&Eq_{\varphi \varphi }=r^{2}\psi ^{\prime \prime
}(r)+2r\psi ^{\prime }(r)+2\Lambda r^{2}  \notag \\
&&-m_{g}^{2}Cc_{1}r-\frac{2q^{2}e^{-\gamma }}{r^{2}},  \label{GR2}
\end{eqnarray}
where $Eq_{tt}$, $Eq_{rr}$, $Eq_{\theta \theta }$ and $Eq_{\varphi \varphi }$
are related to components of $tt$, $rr$, $\theta \theta $\ and $\varphi
\varphi $ of the equations of motion (Eq. (\ref{field eq})).

By considering equations of motion (Eqs. (\ref{GR1}) and (\ref{GR2})), we
can obtain the metric function which leads to 
\begin{equation}
\psi \left( r\right) =1-\dfrac{m_{0}}{r}-\frac{\Lambda }{3}r^{2}+\frac{%
q^{2}e^{-\gamma }}{r^{2}}+m_{g}^{2}C\left( \dfrac{c_{1}r}{2}+c_{2}C\right) ,
\label{solution}
\end{equation}%
where $\Lambda $, is the cosmological constant. $m_{0}$ is a constant that
is related to the total mass of the black hole. Notably, in the absence of
the massive parameters ($c_{1}=c_{2}=0$, or massless graviton, i.e., $%
m_{g}=0 $), the solution (\ref{solution}) turns to the black hole solutions
in Einstein-$\Lambda $-ModMax gravity in the form $\psi \left( r\right) =1-%
\frac{2m_{0}}{r}-\frac{\Lambda }{3}r^{2}+\frac{q^{2}e^{-\gamma }}{r^{2}}$.

Our next step is to examine the geometrical structure of the solutions.
First, we will investigate the existence of essential singularities. The
Ricci and Kretschmann scalars of the solutions are, respectively, 
\begin{eqnarray}
R &=&4\Lambda -\frac{3m_{g}^{2}Cc_{1}}{r}-\frac{2m_{g}^{2}C^{2}c_{2}}{r^{2}},
\\
R_{\alpha \beta \gamma \delta }R^{\alpha \beta \gamma \delta } &=&\frac{%
8\Lambda ^{2}}{3}-\frac{4\Lambda m_{g}^{2}Cc_{1}}{r}+\frac{%
2m_{g}^{2}C^{2}\left( m_{g}^{2}c_{1}^{2}-\frac{4\Lambda c_{2}}{3}\right) }{%
r^{2}}  \notag \\
&&+\frac{4m_{g}^{4}C^{3}c_{1}c_{2}}{r^{3}}+\frac{4m_{g}^{4}C^{4}c_{2}^{2}}{%
r^{4}}  \notag \\
&&-\frac{8m_{g}^{2}C^{2}c_{2}m_{0}+4m_{g}^{2}Cc_{1}q^{2}e^{-\gamma }}{r^{5}}
\notag \\
&&+\frac{12m_{0}^{2}+8m_{g}^{2}C^{2}c_{2}q^{2}e^{-\gamma }}{r^{6}}  \notag \\
&&-\frac{48m_{0}q^{2}e^{-\gamma }}{r^{7}}+\frac{56q^{4}e^{-2\gamma }}{r^{8}}.
\end{eqnarray}

These relations confirm the presence of an essential curvature singularity
at $r=0$. In the limit as $r \longrightarrow \infty$, the Ricci and
Kretschmann scalars yield values of $4\Lambda$ and $\frac{8\Lambda^{2}}{3}$,
respectively. This indicates that for $\Lambda>0$ ($\Lambda<0$), the
asymptotical behavior of the solution is (A)dS. Our finding indicates
although the parameters of ModMax and massive gravity modified the spacetime
but the asymptotcal behavior of spacetime is independent of them.

\begin{figure}[tbph]
\centering
\includegraphics[width=0.48\linewidth]{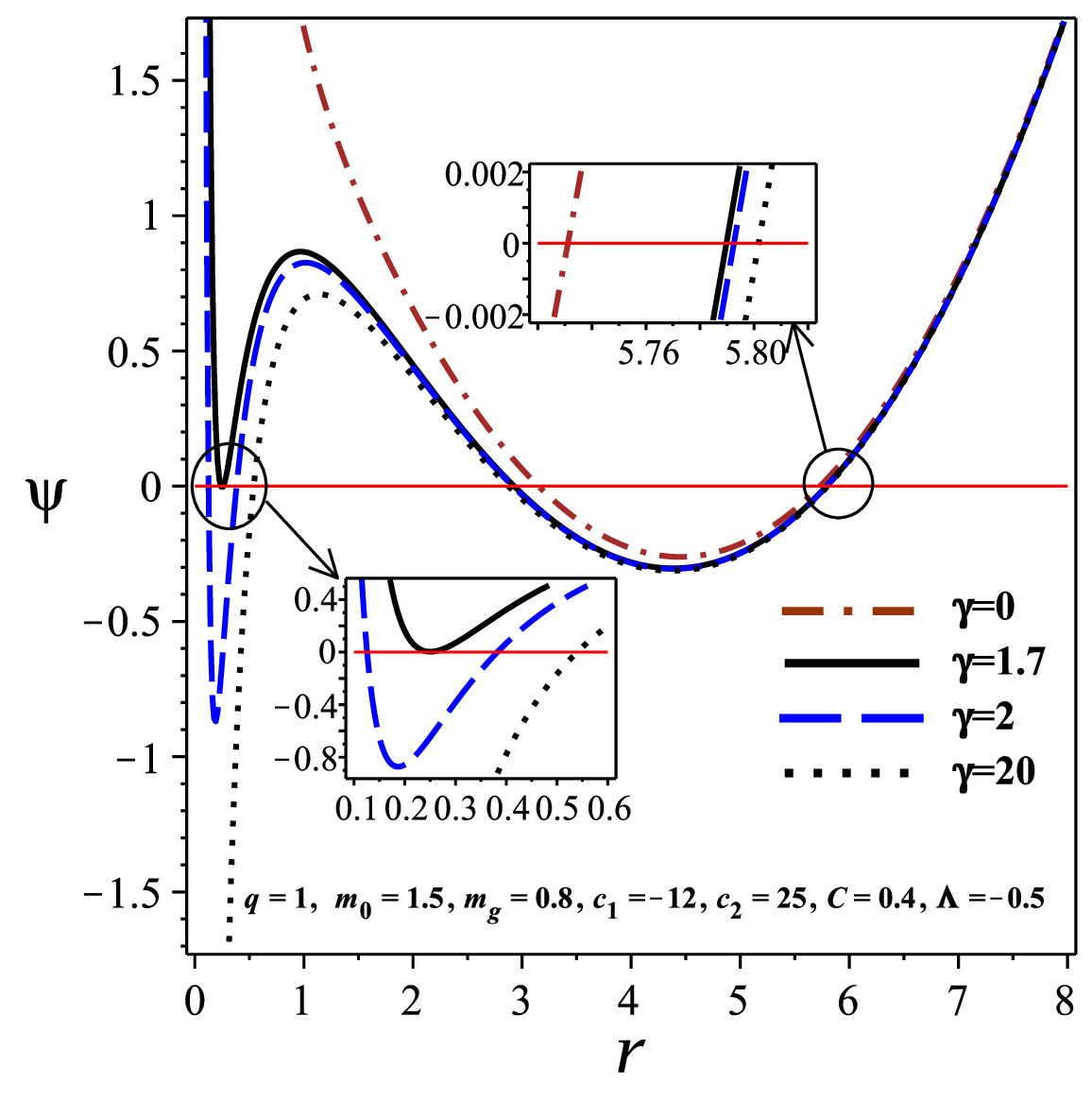} \includegraphics[width=0.48%
\linewidth]{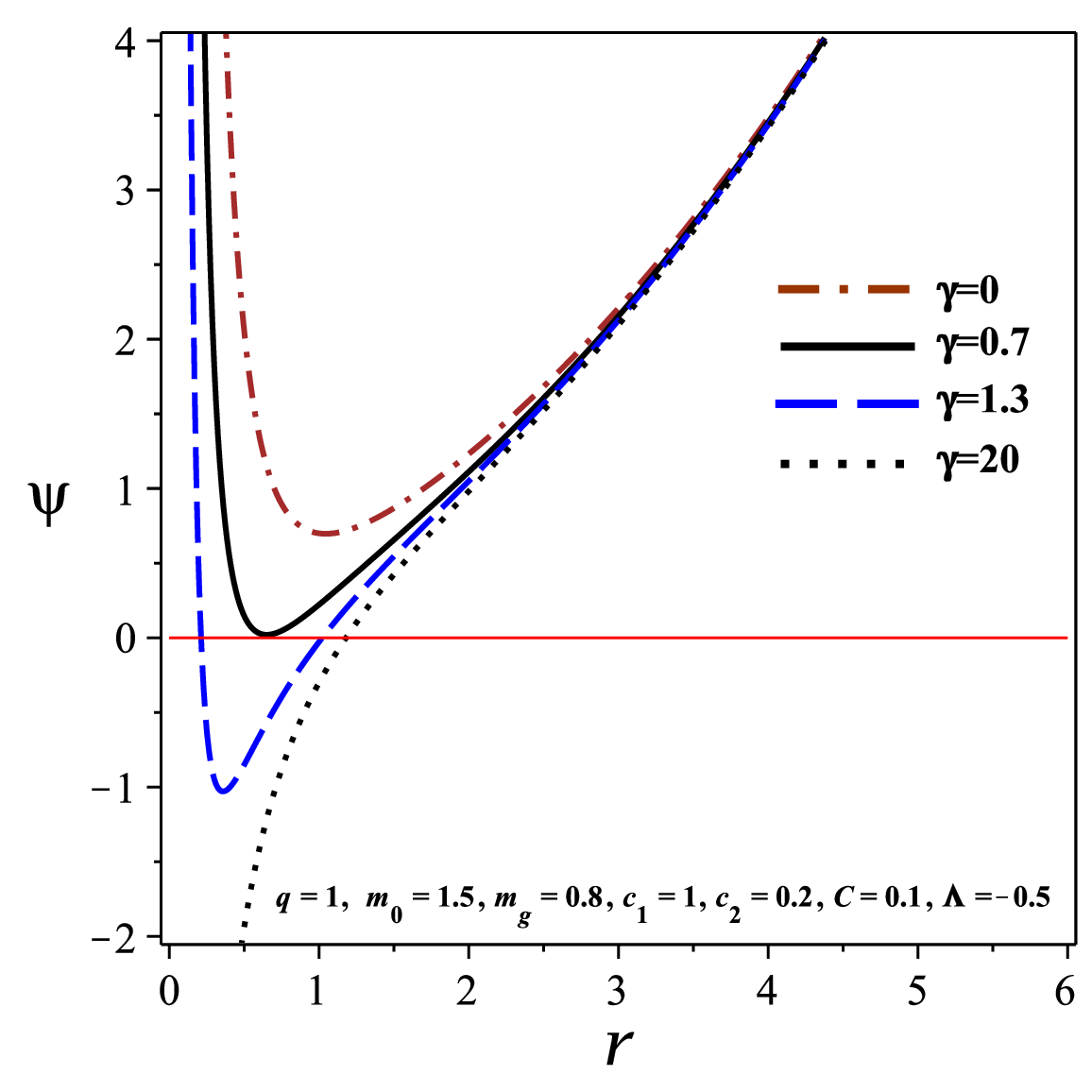} \newline
\caption{$\protect\psi(r)$ (Eq. (\protect\ref{solution})) versus $r$ for
different values of parameters.}
\label{Fig1}
\end{figure}

We plot $\psi (r)$ versus $r$ in Fig. \ref{Fig1} to study the effects of the
ModMax and massive gravity parameters on the number of roots of the solution
(\ref{solution}). We find that these parameters influence the number of
roots of the metric function. Specifically, the number of roots varies with
changes in these parameters. The results are as follows:

i) No real root, indicating a naked singularity.

ii) One real root, representing extremal black holes.

iii) Two real roots, consisting of one inner horizon and one outer horizon
(or event horizon).

iv) Three real roots, including two inner horizons and one event horizon.

Our findings indicate that multiple real roots can exist for a high value of 
$\gamma $ when the parameters of massive gravity are large (see the dashed
line in the left panel of Fig. \ref{Fig1}). Conversely, when the massive
parameters are small, increasing $\gamma$ may lead to the solution (Eq. (\ref%
{solution})) having two real roots, similar to ordinary charged black hole
solutions (see the right panel in Fig. \ref{Fig1}). In
addition, there is a critical value for $\gamma$ (denoted as $\gamma_{\text{%
critical}}$) that determines the number of real roots. Specifically, for $%
\gamma < \gamma_{\text{critical}}$, the number of real roots decreases from
four to two (see the continuous and dotted-dashed lines in the left panel of
Fig. \ref{Fig1}). However, for $\gamma > \gamma_{\text{critical}}$, there
are three real roots (see the dotted line in Fig. \ref{Fig1}).

\section{Thermodynamics Quantities in the non-extended phase space}

To extract the thermodynamic properties of black holes in ModMax-dRGT-like
massive gravity, we first express the geometrical mass ($m_{0}$) in terms of
the radius of the event horizon $r_{+}$, the electrical charge $q$, ModMax's
parameter ($\gamma$), the cosmological constant ($\Lambda $), and the
parameters of massive gravity ($c_{1}$, $c_{2}$, and $C$). To achieve this,
we solve the equation $g_{tt}=\psi(r)=0$, which leads to 
\begin{equation}
m_{0}=r_{+}-\frac{\Lambda r_{+}^{3}}{3}+\frac{q^{2}e^{-\gamma }}{r_{+}}%
+m_{g}^{2}Cr_{+}\left( \dfrac{c_{1}r_{+}}{2}+Cc_{2}\right) .  \label{mm}
\end{equation}

The Hawking temperature is given by $T=\frac{\kappa }{2\pi }$, where $\kappa 
$ is the surface gravity. The surface gravity is defined as $\kappa =\sqrt{%
\frac{-1}{2}\left( \nabla {\mu }\chi {\nu }\right) \left( \nabla ^{\mu }\chi
^{\nu }\right) }$. By using the Killing vector $\chi =\partial _{t}$, and
the spacetime (\ref{metric}), we can express the surface gravity as $\kappa =%
\frac{1}{2}\left. \frac{\partial \psi (r)}{\partial r}\right\vert _{r=r{+}}$%
. Thus, the Hawking temperature of these black holes is given by 
\begin{equation}
T=\frac{1}{4\pi }\left( \frac{1}{r_{+}}-\Lambda r_{+}-\frac{q^{2}e^{-\gamma }%
}{r_{+}^{3}}+m_{g}^{2}C\left( c_{1}+\frac{Cc_{2}}{r_{+}}\right) \right) .
\label{TemII}
\end{equation}

\begin{figure}[tbph]
\centering
\includegraphics[width=0.48\linewidth]{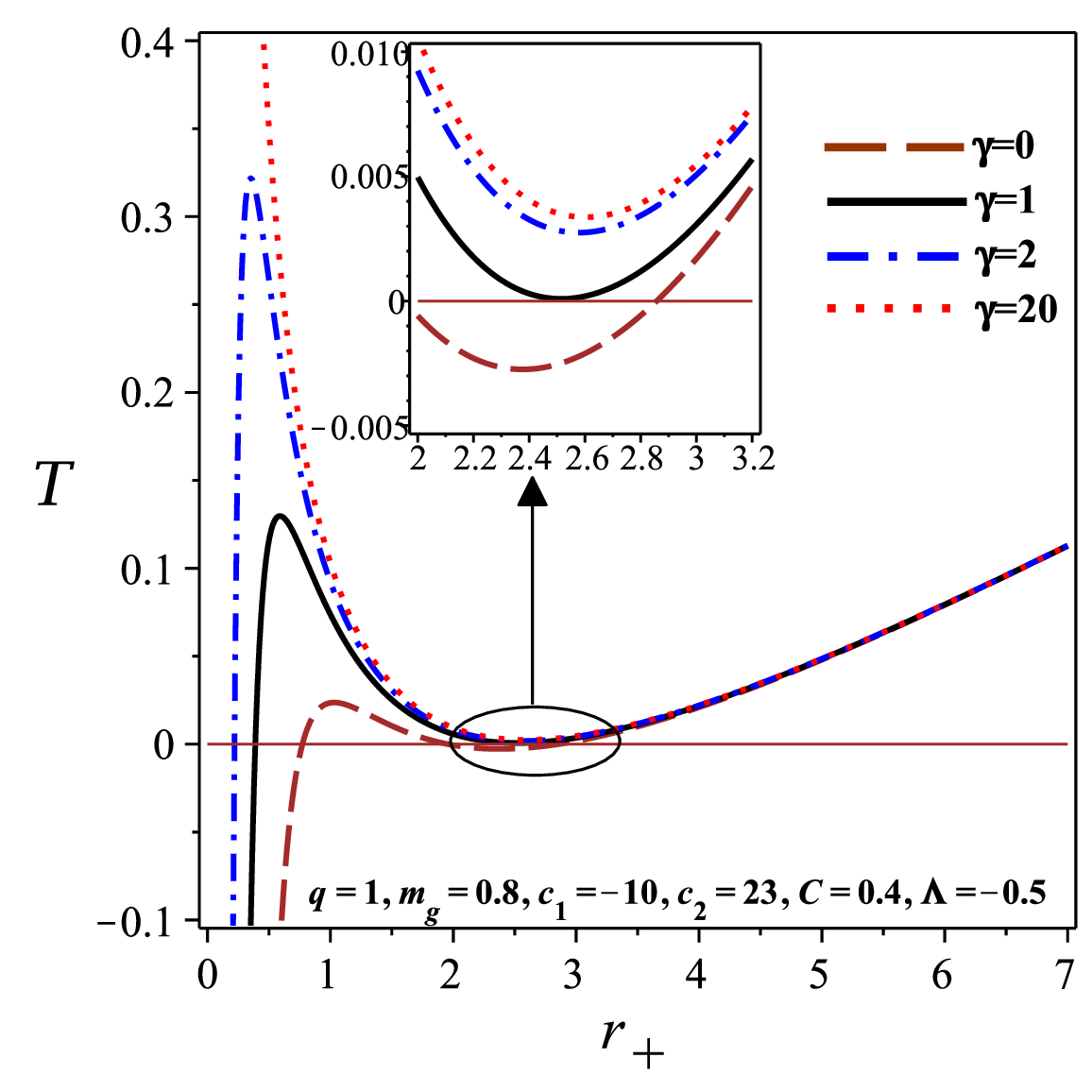} \includegraphics[width=0.48%
\linewidth]{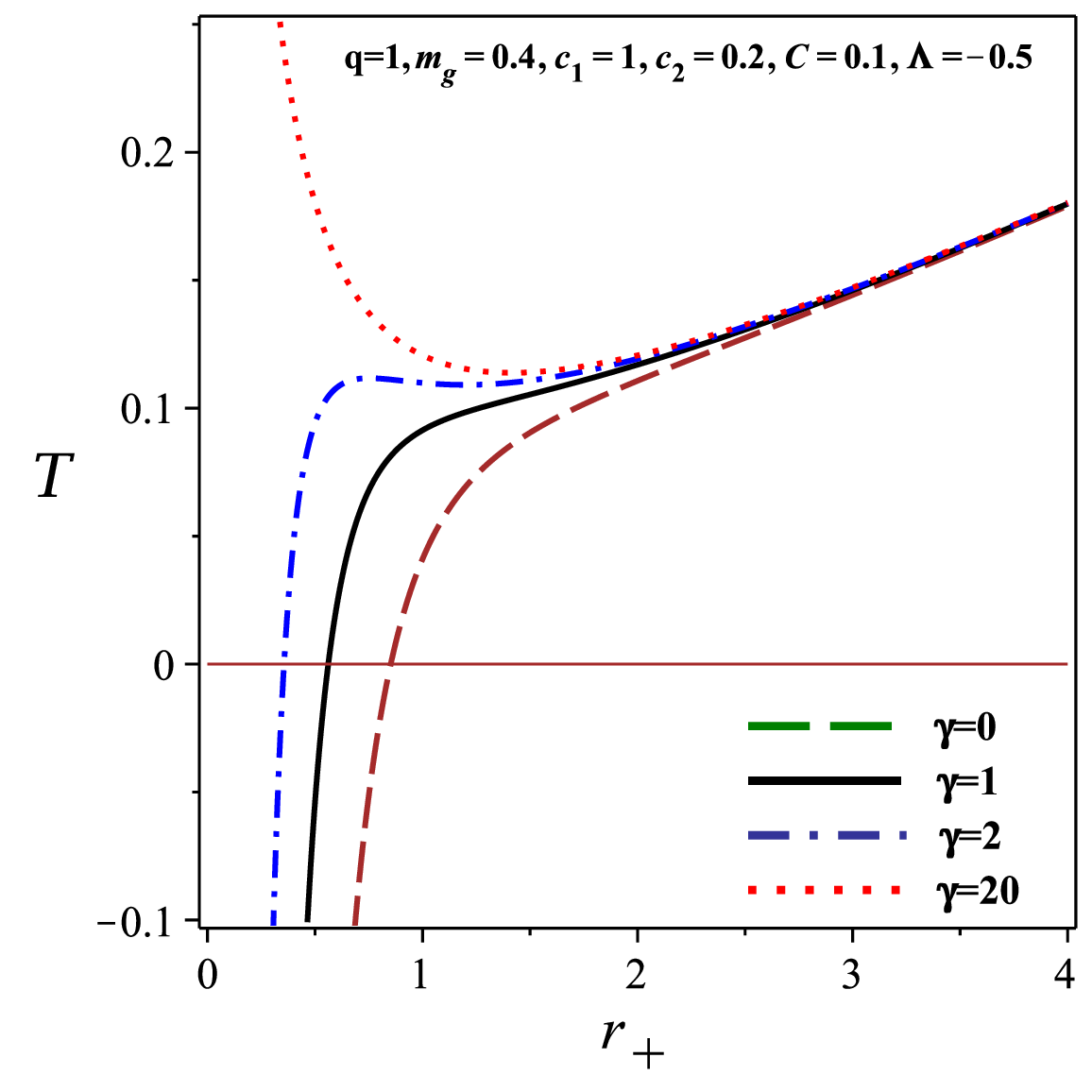} \newline
\caption{Temperature (Eq. (\protect\ref{TemII})) versus $r_{+}$ for
different values of parameters.}
\label{Fig2}
\end{figure}

The parameters of ModMax and the massive theory of gravity clearly alter the
Hawking temperature. Our analysis indicates that:

\textbf{For large values of massive gravity's parameters:}

i) As the ModMax parameter increases, the number of roots of the temperature
decreases. Indeed, the temperature can have one, two, or three roots
depending on the value of $\gamma$ (refer to the dashed, continuous, and
dashed-dotted lines in the left panel of Fig. \ref{Fig2}).

ii) There are no roots for the temperature when $\gamma$ takes on very large
values (see dotted lines in the left panel of Fig. \ref{Fig2}). In this
scenario, the temperature remains positive, indicating that the black holes
are always physical objects.

iii) by considering special values for parameters of massive gravity, the
temperature of medium black holes is positive for the ModMax field and
negative for the Maxwell field. In other words, medium black holes in the
Maxwell-dRGT-like massive gravity (see dashed line in Fig. \ref{Fig2}) are
non- physical systems due to their negative temperatures. In contrast,
medium black holes in the ModMax-dRGT-like massive gravity are physical
systems because their temperatures are positive (see continuous and
dashed-dotted lines in Fig. \ref{Fig2}). This contrasting behavior stems
from the presence of the ModMax field, highlighting a significant difference
in temperature behavior between Maxwell and ModMax fields.

\textbf{For small values of massive gravity's parameters:}

i) There is only one root for the temperature (see the dashed, continuous,
and dashed-dotted lines in the right panel of Fig. \ref{Fig2}). This root
decreases as the ModMax parameter increases, indicating that an increase in
the ModMax parameter leads to a larger physical area.

ii) Similar to the previous case, for very large values of $\gamma$, there
is no root for the temperature (see the dotted lines in the right panel of
Fig. \ref{Fig2}). In other words, these black holes are always physical
objects because their temperatures are positive.

It is noteworthy that large black holes can have a positive temperature when
the cosmological constant $\Lambda$ is negative. Specifically, large Anti de
Sitter (AdS) black holes, characterized by $\Lambda <0$, can have a positive
temperature because the asymptotic limit of the temperature depends on the
cosmological constant (see Fig. \ref{Fig2}). 

We can get the total electric charge of these black holes in the following
form 
\begin{equation}
Q=\frac{1}{4\pi }\int_{\mathcal{H}}\mathcal{F}=qe^{-\gamma },  \label{Q}
\end{equation}%
where $\mathcal{F}$ is the Maxwell invariant. Additionally, $\mathcal{H}$ is
defined as a closed space-like two-surface. As can be observed, the total
electric charge of these black holes depends on the ModMax parameter,
vanishing as $\gamma \rightarrow \infty$ (i.e., $lim_{\gamma \rightarrow
\infty} Q\rightarrow 0$).

The electric potential at the event horizon ($U$) with respect to the
reference ($r\rightarrow \infty $) is obtained as 
\begin{equation}
U=\frac{q}{r_{+}}.  \label{elcpo}
\end{equation}

To extract the entropy of these black holes, one can apply the area law,
which states 
\begin{equation}
S=\frac{\mathcal{A}}{4}=\pi r_{+}^{2},  \label{S}
\end{equation}%
where $\mathcal{A}=\left. \int_{0}^{2\pi }\int_{0}^{\pi }\sqrt{g_{\theta
\theta }g_{\varphi \varphi }}d\theta d\varphi \right\vert _{r=r_{+}}=4\pi
r_{+}^{2}$ is the horizon area.

Using the Ashtekar-Magnon-Das (AMD) approach \cite{AMDI,AMDII}, we can
determine the total mass of these black holes, which is 
\begin{eqnarray}
M &=&\frac{m_{0}}{2}=\frac{r_{+}}{2}-\frac{\Lambda r_{+}^{3}}{6}+\frac{%
q^{2}e^{-\gamma }}{2r_{+}}  \notag \\
&&+\frac{m_{g}^{2}Cr_{+}}{2}\left( \frac{c_{1}r_{+}}{2}+Cc_{2}\right) .
\label{MM}
\end{eqnarray}%
It is important to note that we replace the geometrical mass (\ref{mm}) into 
$M=\frac{m_{0}}{2}$ to derive the total mass (\ref{MM}).

\begin{figure}[tbph]
\centering
\includegraphics[width=0.48\linewidth]{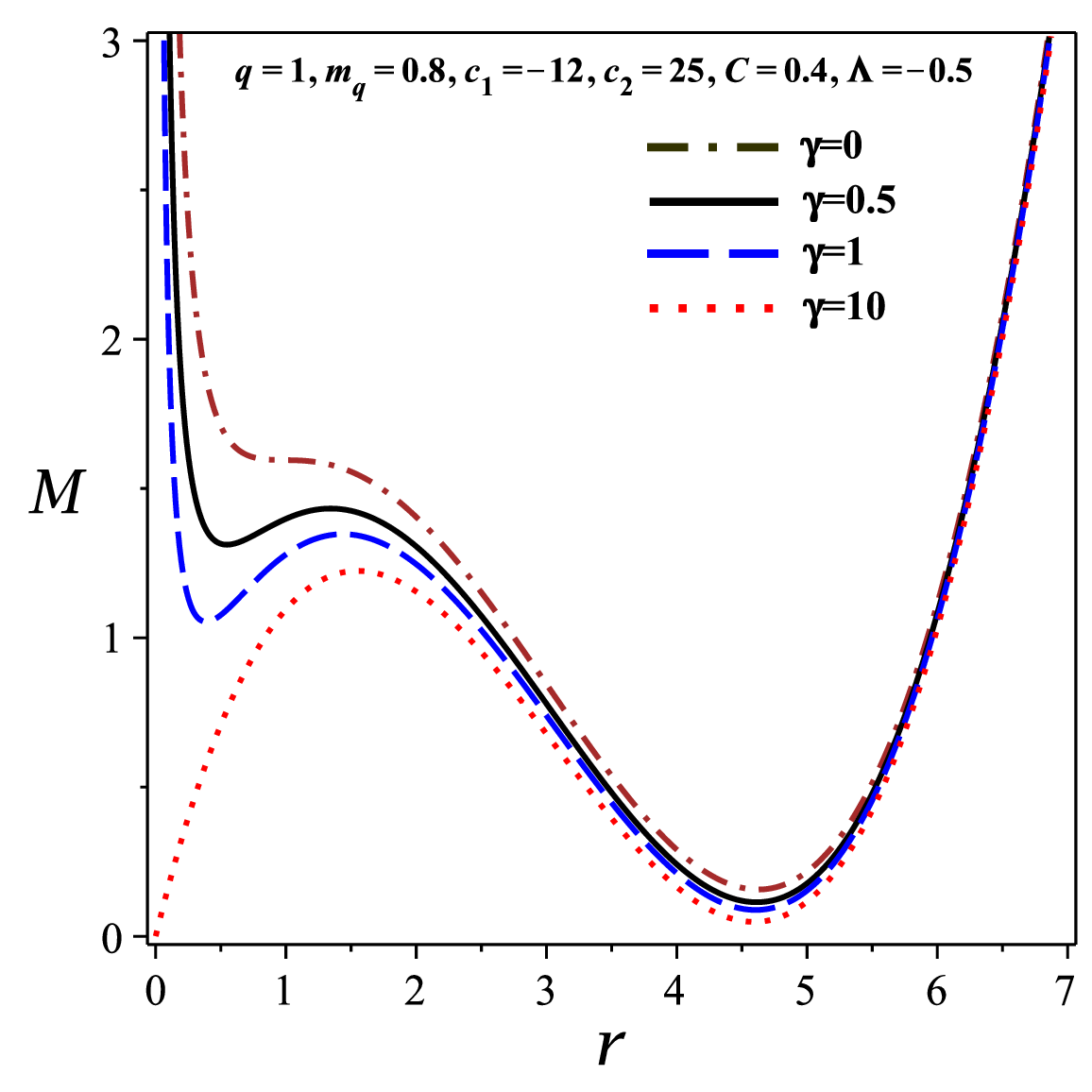} \includegraphics[width=0.48%
\linewidth]{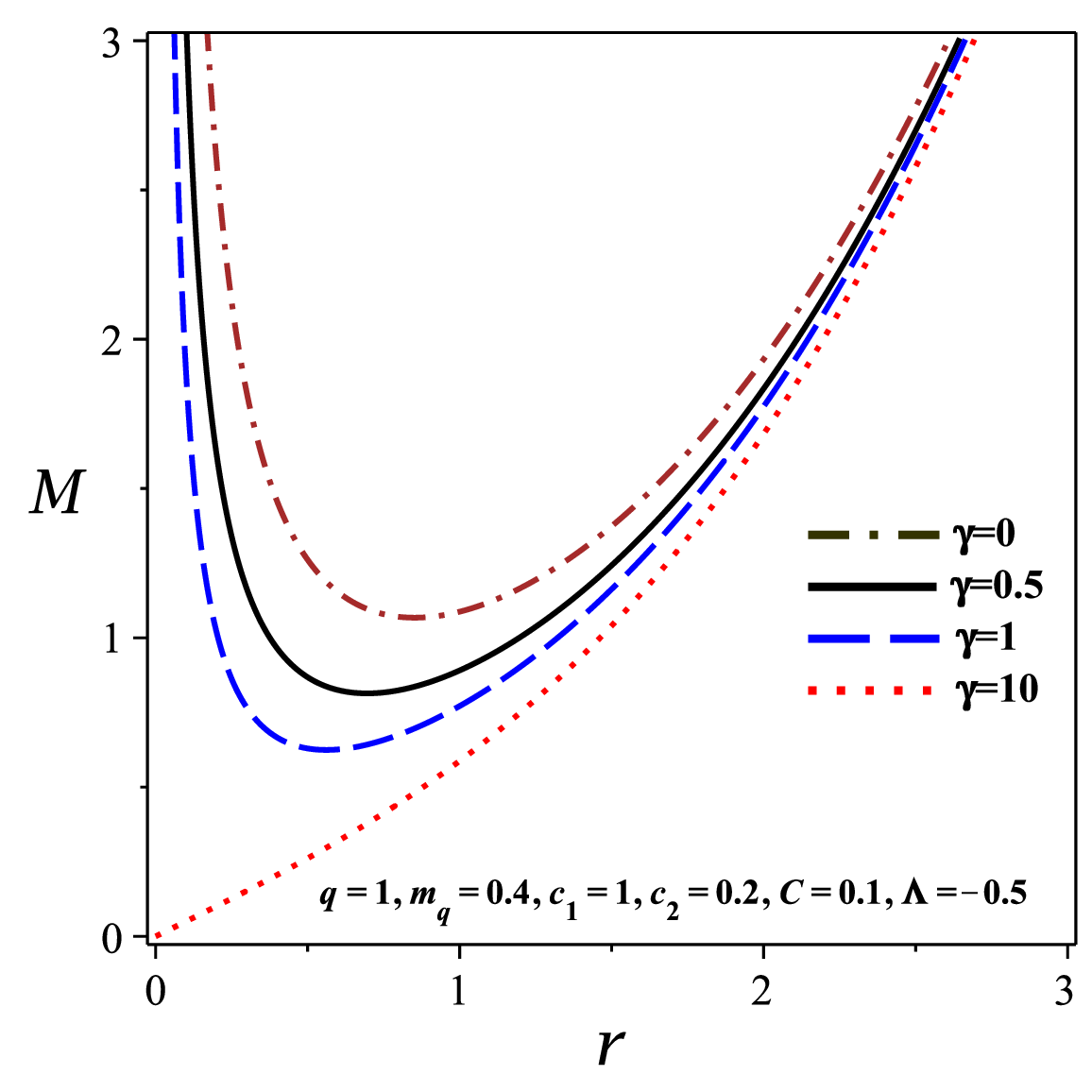} \newline
\caption{Total mass (Eq. (\protect\ref{MM})) versus $r_{+}$ for different
values of parameters.}
\label{Fig3}
\end{figure}

Our analysis indicates that the high-energy limit of the mass (i.e., $%
\underset{r_{+}\rightarrow 0}{\lim M}=\frac{q^{2}e^{-\gamma }}{2r_{+}}$)
depends on the electric charge and the parameter of ModMax. This leads to an
interesting behavior in a high-energy limit of the mass. In other words, the
total mass of black holes can be zero when $\gamma \rightarrow \infty$,
which is due to the existence of the parameter of ModMax theory (for more
details, see dotted line in Fig. \ref{Fig3}).

Our findings reveal that the asymptotic limit of the mass (i.e., $\underset{%
r_{+}\rightarrow \infty }{\lim M}=\frac{r_{+}}{2}-\frac{\Lambda r_{+}^{3}}{6}%
+\frac{m_{g}^{2}Cr_{+}}{2}\left( \frac{c_{1}r_{+}}{2}+Cc_{2}\right) $)
depends on the cosmological constant, and the parameters of dRGT-like
massive gravity.

In addition, for large values of massive gravity's parameters, there are
three critical points (two minima and one maximum) for total mass (see the
left panel in Fig. \ref{Fig3}, excluding the dotted line). However, for
small values of massive gravity's parameters, the total mass of these black
holes reaches a minimum point (see the right panel in Fig. \ref{Fig3},
excluding the dotted line). As mentioned, for both large and small values of
the parameters of massive gravity, the total mass of black holes is zero for very large $\gamma$ (see the dotted line in Fig. \ref{Fig3}). 

By comparing the obtained results of the Hawking temperature and total mass
at the asymptotic limit of these black holes, we find that large AdS black
holes in ModMax-dRGT-like massive gravity can be supermassive black holes
(or big black holes), as both the temperature and mass of large black holes
are positive when $\Lambda <0$.

\subsection{Local Stability}
To study the local stability of charged black holes in the ModMax-dRGT-like massive gravity theory, we
calculate their heat capacity. In the context of the canonical ensemble, the
local stability of a thermodynamic system, such as a black hole, can be
assessed through its heat capacity.

The heat capacity is defined as 
\begin{equation}
C_{q}=\frac{T}{\left( \frac{\partial T}{\partial S}\right) _{q}}=T\left( 
\frac{\frac{\partial S}{\partial r_{+}}}{\frac{\partial T}{\partial r_{+}}}%
\right) _{q},  \label{Heat}
\end{equation}%
by considering Eqs. (\ref{TemII}), and (\ref{S}), we obtain the heat
capacity in the following form 
\begin{equation}
C_{q}=\frac{2\pi r_{+}^{2}\left( 1-\Lambda r_{+}^{2}-\frac{q^{2}e^{-\gamma }%
}{r_{+}^{2}}+m_{g}^{2}C\left( c_{1}r_{+}+Cc_{2}\right) \right) }{\frac{%
3q^{2}e^{-\gamma }}{r_{+}^{2}}-1-\Lambda r_{+}^{2}-m_{g}^{2}C^{2}c_{2}},
\label{C}
\end{equation}

The heat capacity offers insights into physical limitations and phase
transition points within the context of the canonical ensemble. Specifically:

i) The roots of the heat capacity ($C_{q} = 0$) indicate physical limitation
points, as they delineate the boundary between physical black holes (where $%
T>0$) and non-physical black holes (where $T<0$). At these limitation
points, the system experiences a change in the sign of the heat capacity.

ii) The divergences in heat capacity signify the critical points of phase
transitions for black holes, denoted by $\left( \frac{\partial T}{\partial S}
\right)_{q} = 0$.

\begin{figure}[tbph]
\centering
\includegraphics[width=0.48\linewidth]{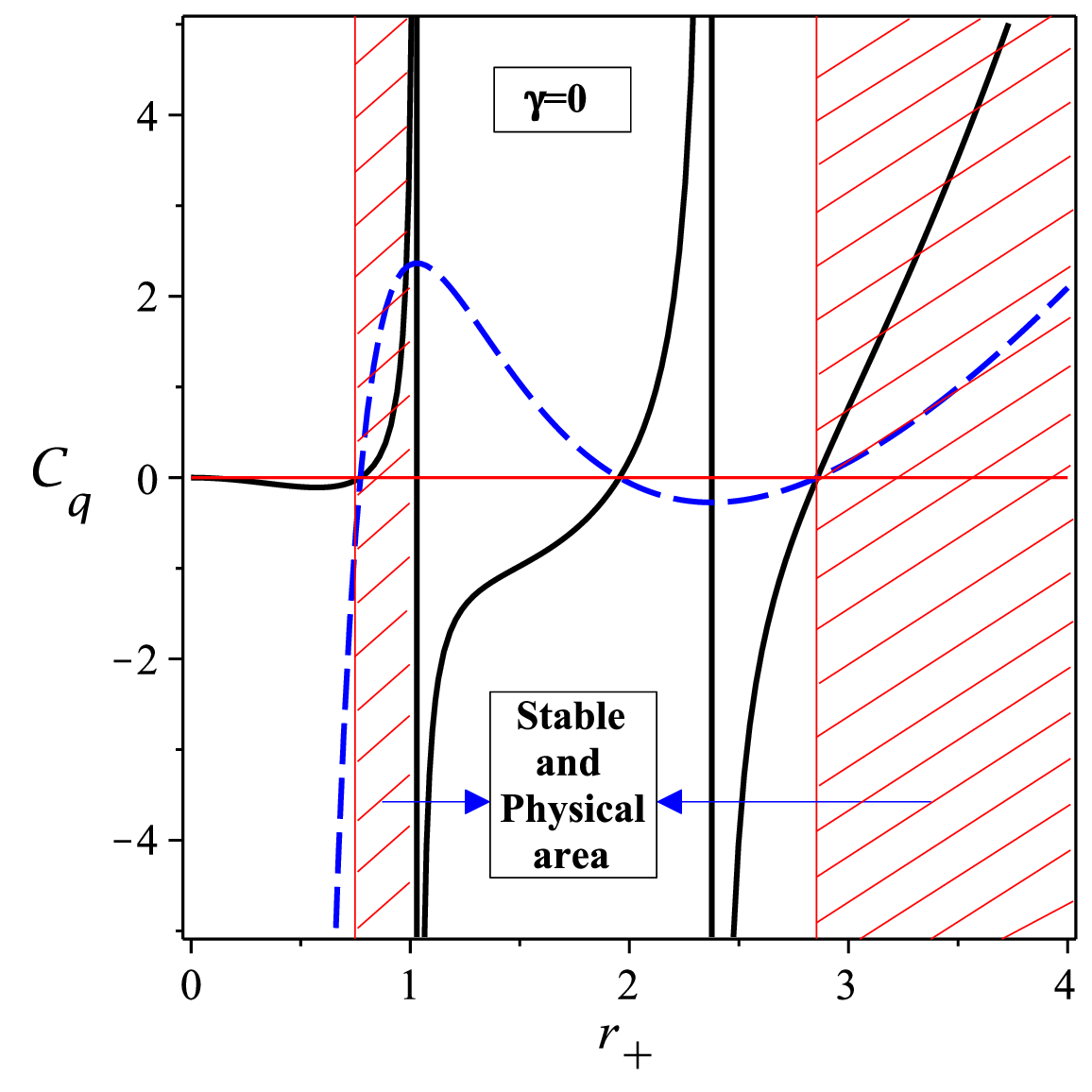} \includegraphics[width=0.48%
\linewidth]{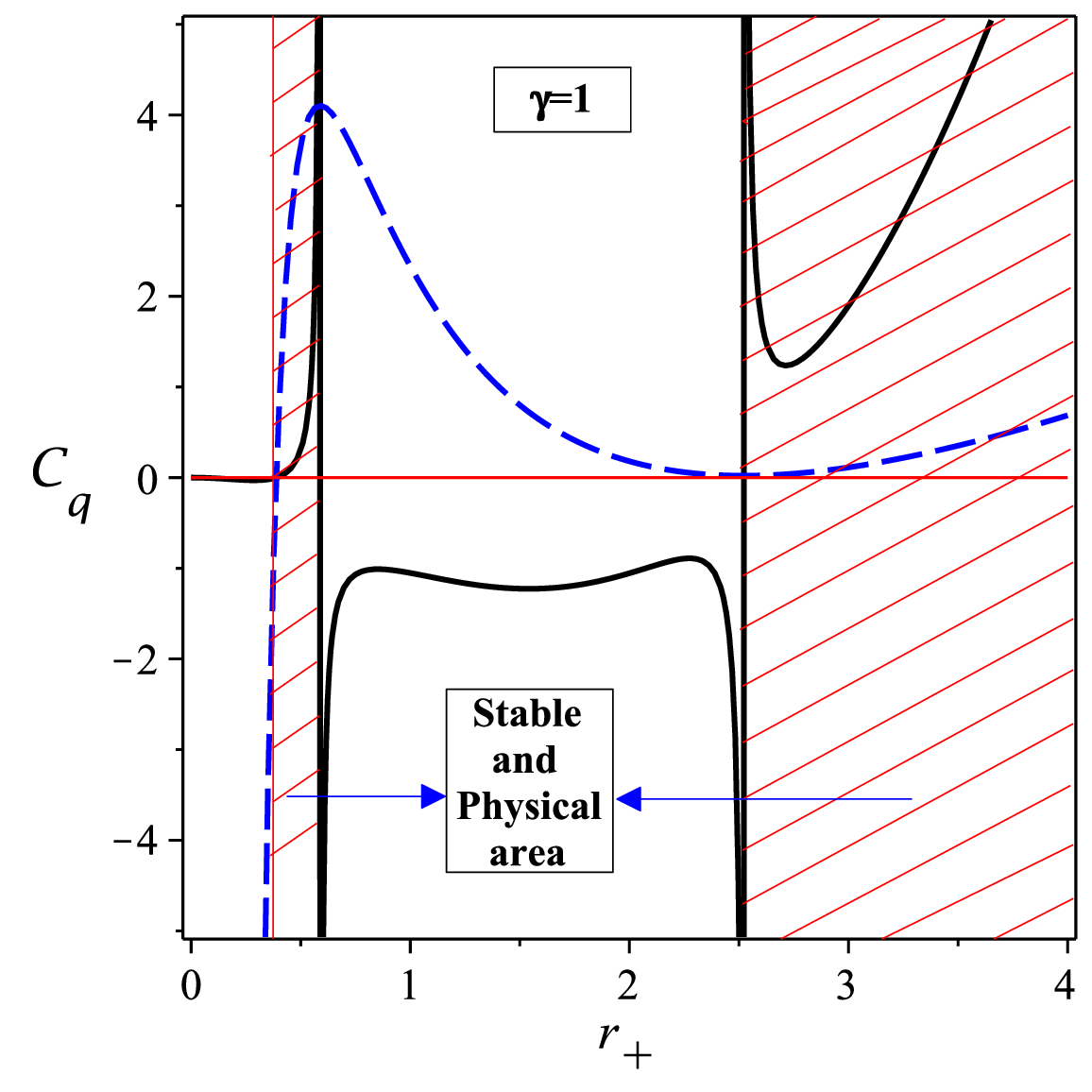} \newline
\includegraphics[width=0.48\linewidth]{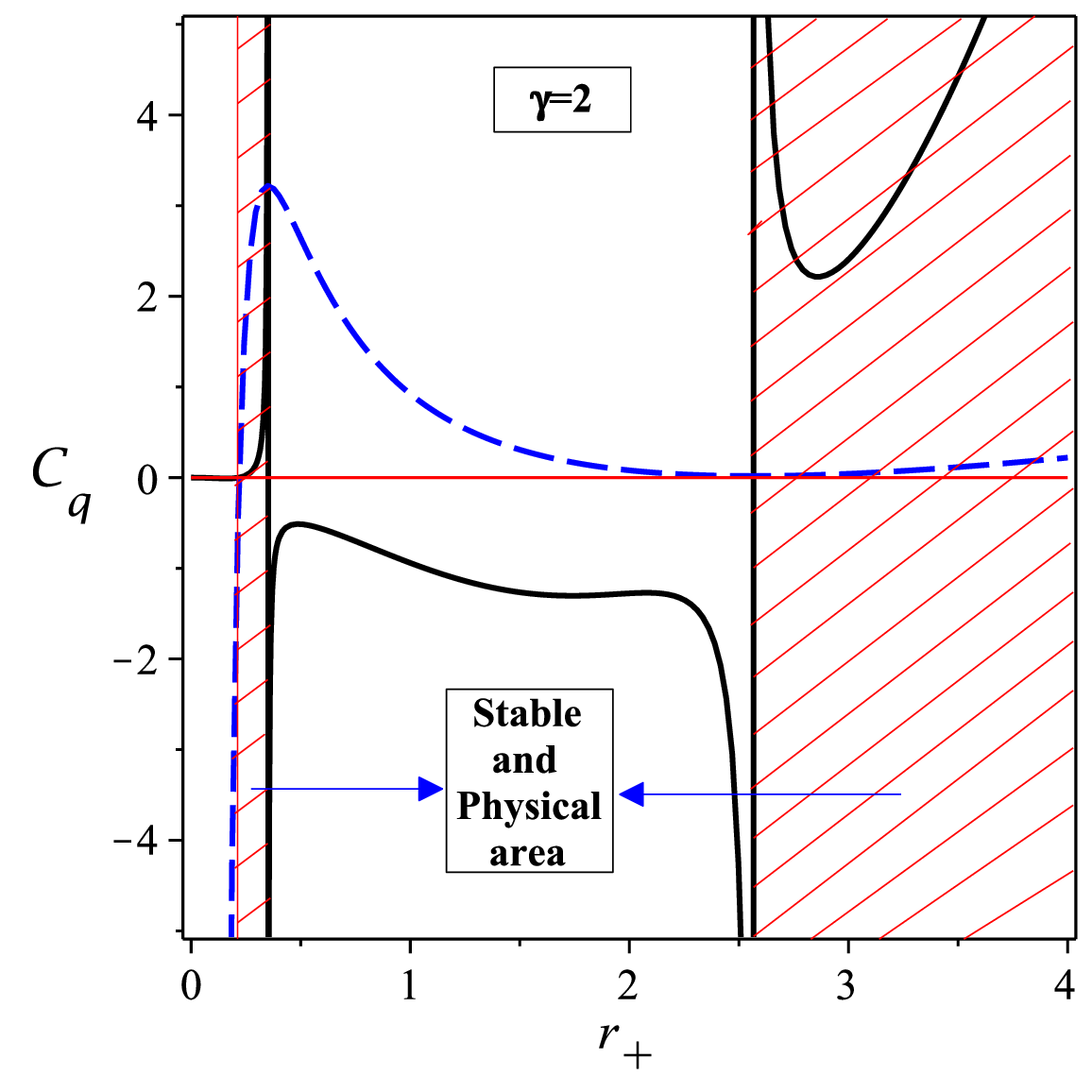} \includegraphics[width=0.48%
\linewidth]{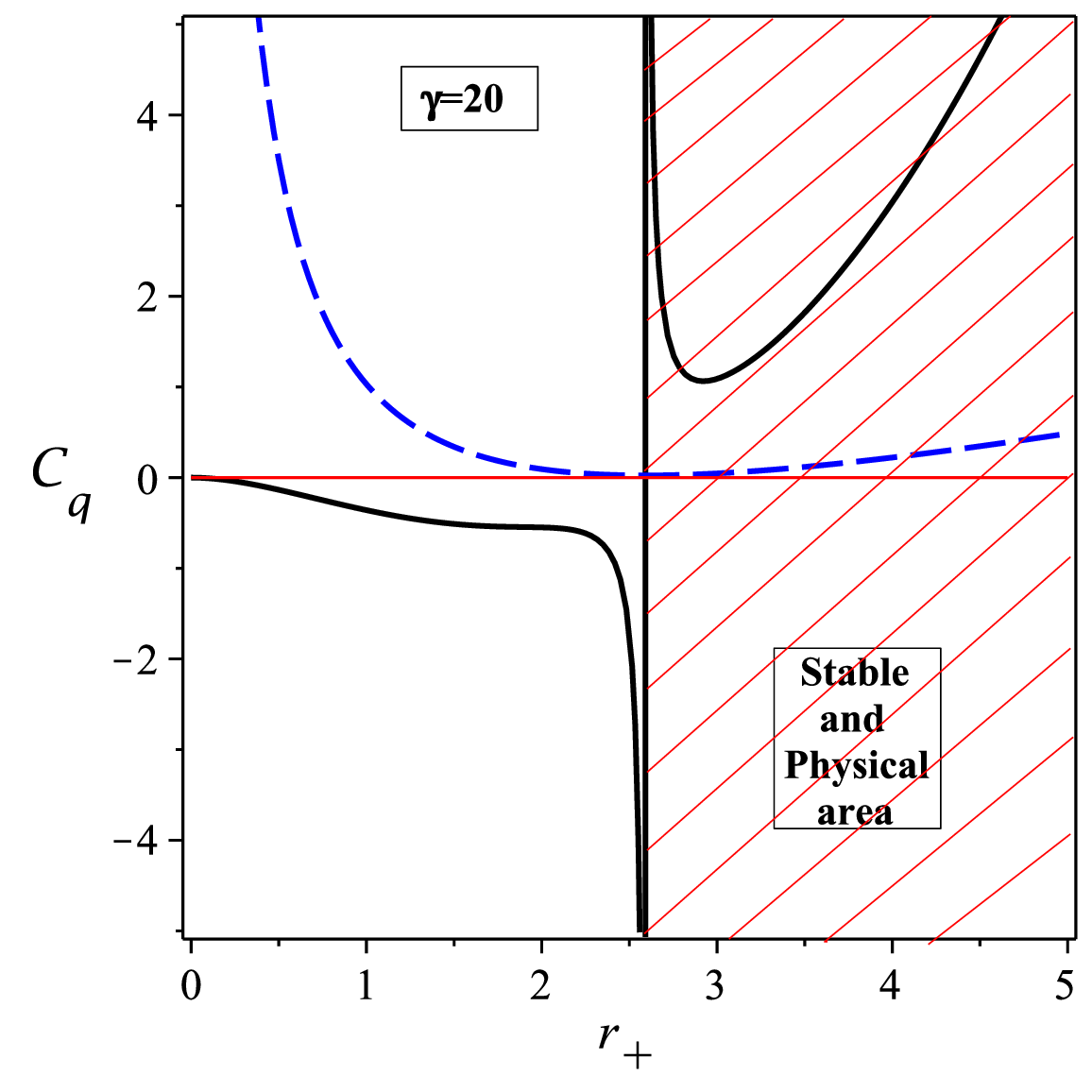} \newline
\caption{Heat capacity $C_{q}$ (Eq. (\protect\ref{C})) versus $%
r_{+}$ for $q=1$, $m_{g}=0.8$, $\Lambda=-0.5$ and large values of massive
parameters which are $c=0.4$, $c_{1}=-10$, $c_{2}=23$.}
\label{Fig4}
\end{figure}

\begin{figure}[tbph]
\centering
\includegraphics[width=0.48\linewidth]{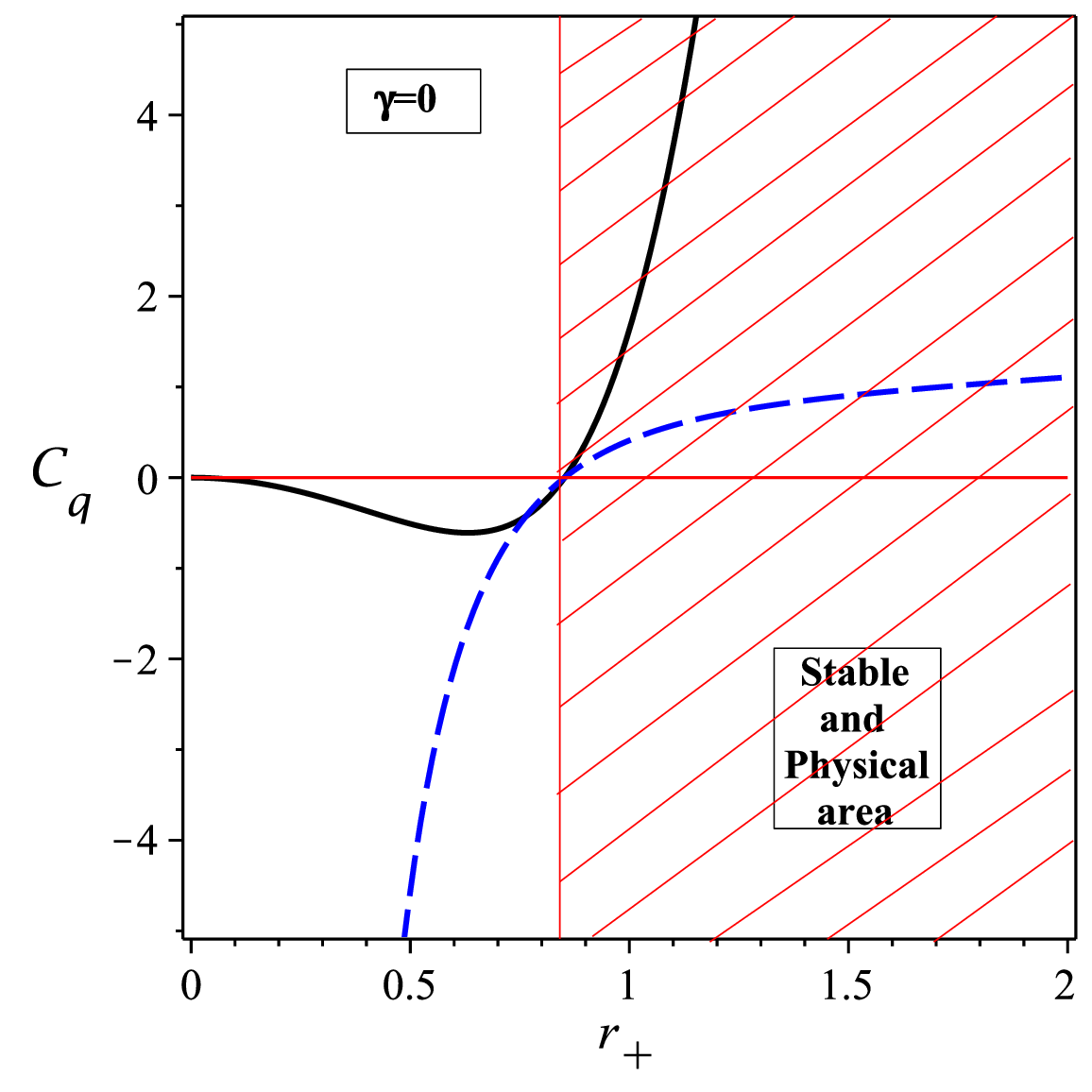} \includegraphics[width=0.48%
\linewidth]{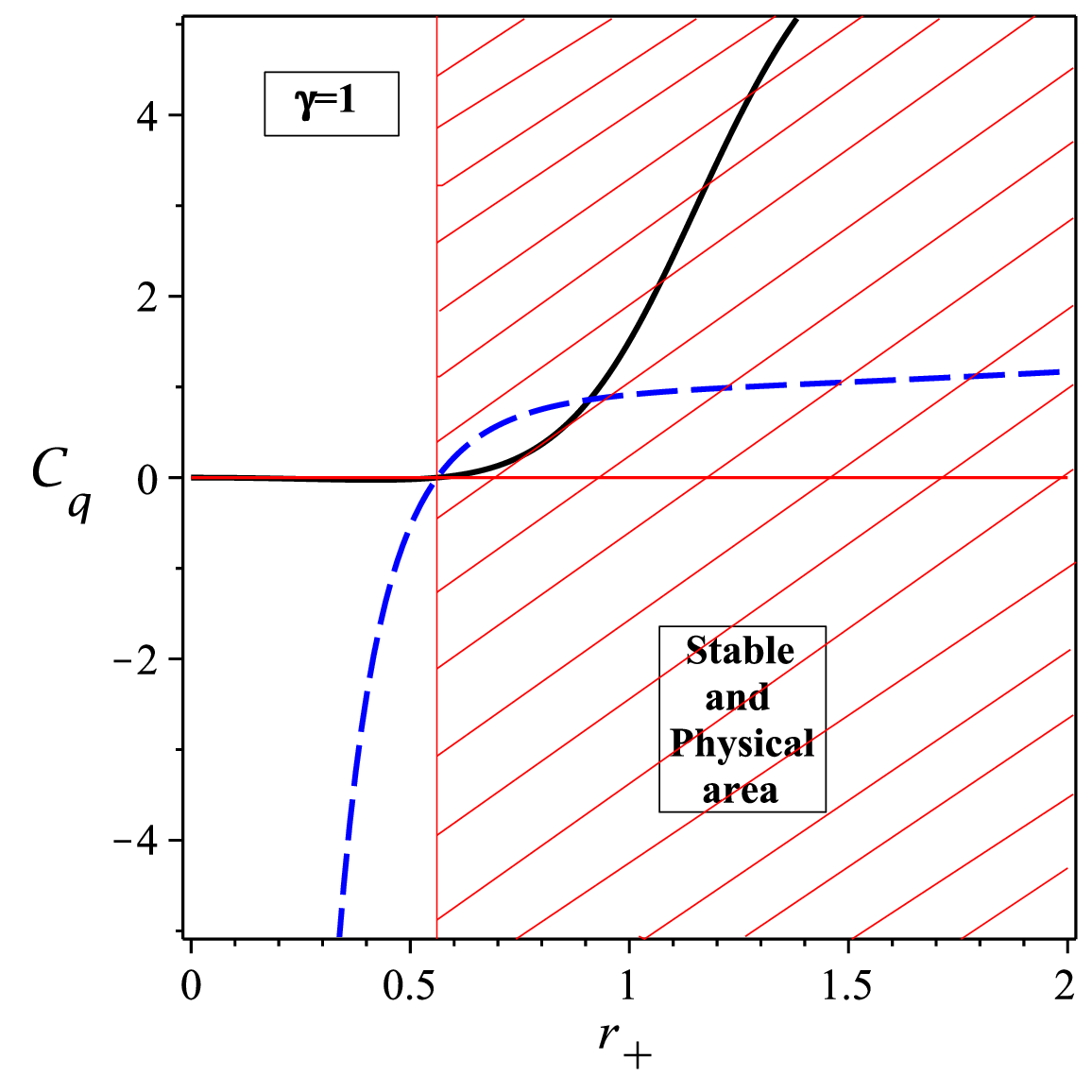} \newline
\includegraphics[width=0.48\linewidth]{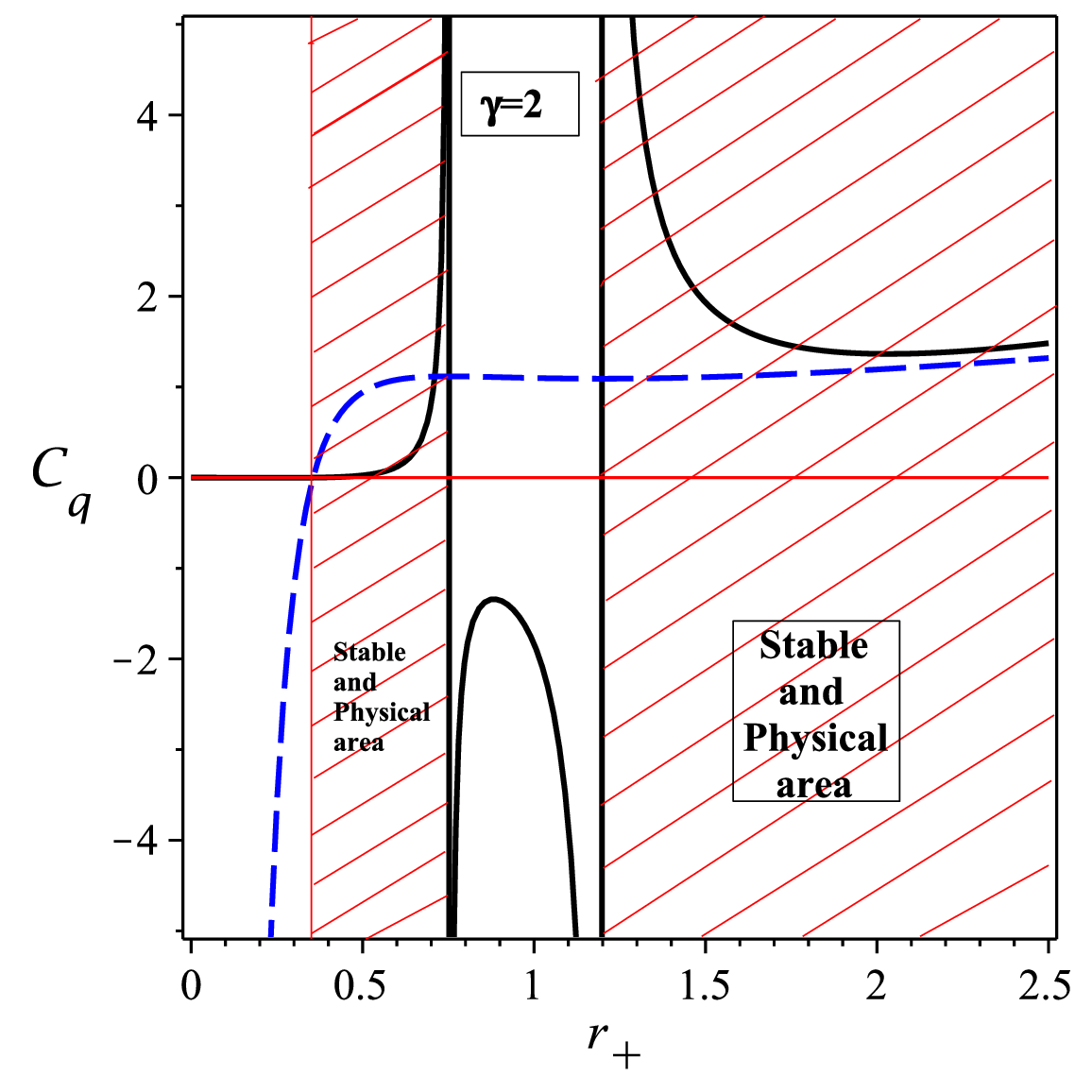} \includegraphics[width=0.48%
\linewidth]{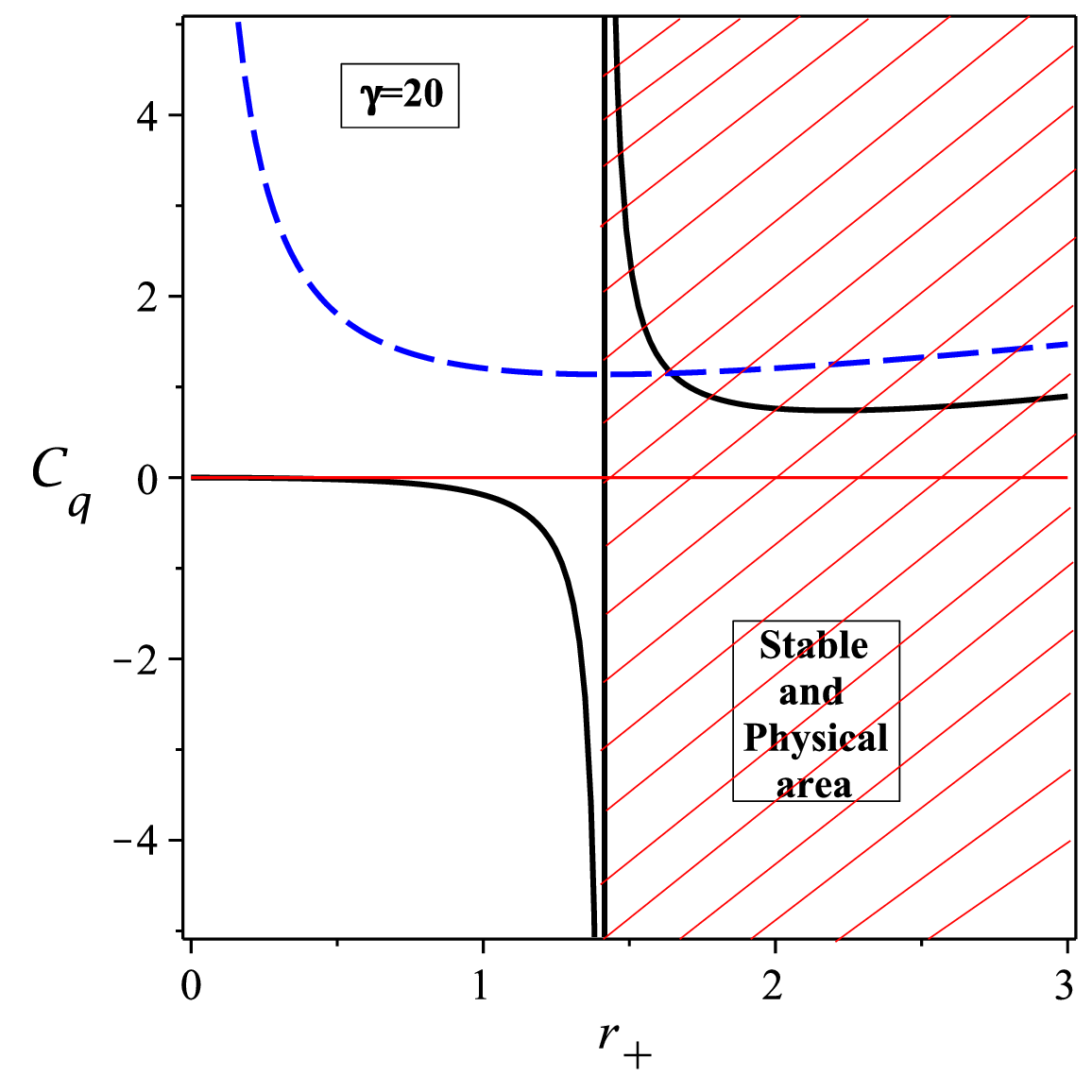} \newline
\caption{Heat capacity $C_{q}$ (Eq. (\protect\ref{C})) versus $%
r_{+}$ for $q=1$, $m_{g}=0.4$, $\Lambda =-0.5$ and small values of massive
parameters which are $c=0.1$, $c_{1}=1$, $c_{2}=0.2$.}
\label{Fig5}
\end{figure}

To determine the phase transition critical and limitation points of these
black holes in relation to heat capacity, we present our findings in Figs. %
\ref{Fig4} and \ref{Fig5}. Our results indicate the following:

1- For large values of the massive gravity parameters: when $\gamma = 0$,
there are two stable and physical regions where both temperature and heat
capacity are positive (see the upper-left panel in Fig. \ref{Fig4}). This
means that medium and large black holes are stable and physical systems.
However, as $\gamma$ increases, the stable region for large black holes
expands while it decreases for medium black holes (see the upper-right and
lower-left panels in Fig. \ref{Fig4}). Furthermore, at high values of $%
\gamma $, only large black holes remain stable and physical (see the
lower-right panel in Fig. \ref{Fig4}). Notably, increasing the ModMax
parameter reduces the number of phase transition points from two to one.
Additionally, as $\gamma$ increases, the physical limitation points decrease
from three to one.

2- For small values of the massive gravity parameters, only large black
holes are stable and physical when $\gamma = 0$ (see the upper-left panel in
Fig. \ref{Fig5}). As $\gamma$ increases, the stable area expands (see the
upper-right panel in Fig. \ref{Fig5}). For even larger values of $\gamma$,
there are two stable and physical regions (medium and large black holes)
(see the lower-left panel in Fig. \ref{Fig5}). Additionally, at very high
values of $\gamma$, we only observe a single stable and physical region (see
the lower-right panel in Fig. \ref{Fig5}). As observed, there are no phase
transition points for small values of $\gamma$. However, as $\gamma$
increases, phase transition points emerge. Initially, with increasing ModMax
parameter, there are two phase transition points, which reduce to one at
very high values of $\gamma$. Additionally, while there is one physical
limitation point for small values of $\gamma$, there are none for very large
values of $\gamma$.

\subsection{First Law of Thermodynamics}
The obtained conserved and thermodynamics quantities satisfy the first law of thermodynamics as 
\begin{equation}
dM=TdS+UdQ,
\end{equation}%
where $T=\left( \frac{\partial M}{\partial S}\right) _{Q}$, and $U=\left( 
\frac{\partial M}{\partial Q}\right) _{S}$ are, respectively, the Hawking
temperature and the electric potential.

In Ref. \cite{MaZhao}, it is demonstrated that a superficial
inconsistency exists between the conventional first law of black hole
thermodynamics and the Bekenstein-Hawking area law for three types of
regular black holes. In this study, we aim to investigate whether a similar
inconsistency applies to charged black holes in ModMax-dRGT-like massive
gravity. To do this, we calculate the temperature of these black holes using
the first law of thermodynamics ($dM=TdS+UdQ$), which results in 
\begin{eqnarray}
T &=&\left( \frac{\partial M}{\partial S}\right) _{Q}=\left( \frac{\frac{%
\partial M}{\partial r_{+}}}{\frac{\partial S}{\partial r_{+}}}\right) _{Q} 
\notag \\
&=&\frac{\frac{1}{r_{+}}-\Lambda r_{+}-\frac{q^{2}e^{-\gamma }}{r_{+}^{3}}%
+m_{g}^{2}C\left( c_{1}+\frac{Cc_{2}}{r_{+}}\right) }{4\pi },  \label{Tlaw}
\end{eqnarray}%
where we consider Eqs. (\ref{S}) and (\ref{MM}) in order to obtain $\frac{%
\partial M}{\partial r_{+}}$ and $\frac{\partial S}{\partial r_{+}}$, which
are 
\begin{eqnarray}
\frac{\partial M}{\partial r_{+}} &=&\frac{1-\Lambda r_{+}^{2}-\frac{%
q^{2}e^{-\gamma }}{r_{+}^{2}}+m_{g}^{2}C\left( c_{1}r_{+}+Cc_{2}\right) }{2},
\\
&&  \notag \\
\frac{\partial S}{\partial r_{+}} &=&2\pi r_{+}.
\end{eqnarray}%
Our analysis indicates that the temperature obtained from Eq. (\ref{Tlaw})
is consistent with the Hawking temperature derived from surface gravity in
Eq. (\ref{TemII}).

Additionally, we can derive the entropy by applying the first law of black
hole thermodynamics in the following form 
\begin{equation}
S=\int \frac{\partial M}{T}=\pi r_{+}^{2}.
\end{equation}%
This is consistent with the entropy obtained from the area law (see Eq. (\ref%
{S})). This indicates that there is no inconsistency between the area law
and the first law of black hole thermodynamics in this theory of gravity
within a non-extended phase space.

\section{Thermodynamics Quantities in the extended phase space}
We expand our study to calculate the various conserved and thermodynamic quantities of
these solutions in the extended phase space. To achieve this, we consider an
analogy between thermodynamic pressure ($P$) and the negative cosmological
constant ($\Lambda $), which is introduced in the following form \cite%
{LambdaPI,LambdaPII} 
\begin{equation}
P=\frac{-\Lambda }{8\pi }.  \label{P}
\end{equation}

Using Eq. (\ref{P}), we rewrite the Hawking temperature (\ref{TemII}) and
the total mass (\ref{MM}) in the extended phase space as 
\begin{eqnarray}
T &=&2r_{+}P+\frac{1}{4\pi }\left( \frac{1}{r_{+}}-\frac{q^{2}e^{-\gamma }}{%
r_{+}^{3}}\right)  \notag \\
&&+\frac{m_{g}^{2}C}{4\pi }\left( c_{1}+\frac{Cc_{2}}{r_{+}}\right) ,
\label{T2} \\
M &=&\frac{r_{+}}{2}+\frac{4\pi r_{+}^{3}P}{3}+\frac{q^{2}e^{-\gamma }}{%
2r_{+}}  \notag \\
&&+\frac{m_{g}^{2}Cr_{+}}{2}\left( \frac{c_{1}r_{+}}{2}+Cc_{2}\right) .
\label{M0}
\end{eqnarray}

The electric charge, electric potential, and entropy in the extended phase
space of black hole solutions in the theory of ModMax dRGT-like massive
gravity are given by%
\begin{eqnarray}
Q &=&qe^{-\gamma },  \label{QE} \\
U &=&\frac{q}{r_{+}},  \label{UE} \\
S &=&\pi r_{+}^{2}.  \label{SE}
\end{eqnarray}%
These quantities reveal that they do not change in the extended phase space.

\subsection{First Law of the Thermodynamics}
By examining the conserved
thermodynamic quantities in the extended phase space, we can express the
extended total mass as 
\begin{eqnarray}
M\left( S,Q,P,c_{1},c_{2}\right) &=&\frac{1+\frac{8}{3}PS+\frac{\pi
Q^{2}e^{\gamma }}{S}}{\frac{2}{\sqrt{\frac{S}{\pi }}}}  \notag \\
&&+\frac{m_{g}^{2}C\left( \frac{c_{1}\sqrt{\frac{S}{\pi }}}{2}+Cc_{2}\right) 
}{\frac{2}{\sqrt{\frac{S}{\pi }}}},  \label{ME}
\end{eqnarray}%
Using the extended mass relation (Eq. (\ref{ME})), we can formulate the
first law of thermodynamics within the context of the extended phase, which
is 
\begin{equation}
dM=TdS+UdQ+VdP+\mathcal{C}_{1}dc_{1}+\mathcal{C}_{2}dc_{2},  \label{law1b}
\end{equation}%
where the conjugate quantities associated with the intensive parameters $S$, 
$Q$, and $P$ are 
\begin{eqnarray}
T &=&\left. \left( \frac{\partial M\left( S,Q,P,c_{1},c_{2}\right) }{%
\partial S}\right) \right\vert _{Q,P,c_{1},c_{2}}  \notag \\
&=&\frac{1+8PS-\frac{\pi Q^{2}e^{\gamma }}{S}+m_{g}^{2}C\left( c_{1}\sqrt{%
\frac{S}{\pi }}+c_{2}C\right) }{4\pi \sqrt{\frac{S}{\pi }}}  \notag \\
&=&\frac{1+8\pi r_{+}^{2}P-\frac{q^{2}e^{-\gamma }}{r_{+}^{2}}%
+m_{g}^{2}C\left( c_{1}r_{+}+Cc_{2}\right) }{4\pi r_{+}},  \label{TS}
\end{eqnarray}
\begin{eqnarray}
U &=&\left. \left( \frac{\partial M\left( S,Q,P,c_{1},c_{2}\right) }{%
\partial Q}\right) \right\vert _{S,P,c_{1},c_{2}}=\frac{Qe^{\gamma }}{\sqrt{%
\frac{S}{\pi }}}  \notag \\
&=&\frac{q}{r_{+}},  \label{US}
\end{eqnarray}
\begin{eqnarray}
V &=&\left. \left( \frac{\partial M\left( S,Q,P,c_{1},c_{2}\right) }{%
\partial P}\right) \right\vert _{S,Q,c_{1},c_{2}}=\frac{4S^{3/2}}{3\sqrt{\pi 
}}  \notag \\
&=&\frac{4\pi r_{+}^{3}}{3}.  \label{VS}
\end{eqnarray}
\begin{eqnarray}
C_{1} &=&\left. \left( \frac{\partial M\left( S,Q,P,c_{1},c_{2}\right) }{%
\partial c_{1}}\right) \right\vert _{S,Q,P,c_{2}}=\frac{m_{g}^{2}CS}{4\pi } 
\notag \\
&=&\frac{m_{g}^{2}Cr_{+}^{2}}{4},  \label{C1S}
\end{eqnarray}
\begin{eqnarray}
C_{2} &=&\left. \left( \frac{\partial M\left( S,Q,P,c_{1},c_{2}\right) }{%
\partial c_{2}}\right) \right\vert _{S,Q,P,c_{1}}=\frac{m_{g}^{2}C^{2}\sqrt{%
\frac{S}{\pi }}}{2}  \notag \\
&=&\frac{m_{g}^{2}C^{2}r_{+}}{2},  \label{C2S}
\end{eqnarray}

\subsection{Smarr Relation }
By considering Eqs. (\ref{T2}), (\ref{M0}), (\ref{QE}), (\ref{UE}), (\ref{SE}), and (\ref{VS}), we can derive the Smarr
relation, which is 
\begin{equation}
M=2TS+UQ-2PV-c_{1}C_{1}.  \label{smarr}
\end{equation}

\subsection{Isoperimetric Ratio}
The isoperimetric ratio is defined as 
\begin{equation}
R=\left( \frac{3V}{4\pi }\right) ^{1/3}\left( \frac{4\pi }{\mathcal{A}}%
\right) ^{1/2},  \label{R}
\end{equation}%
by replacing the horizon area $\mathcal{A}$ (Eq. (\ref{S})) with the
thermodynamic volume $V$ (Eq. (\ref{VS})) in the equation (\ref{R}), we find
that $R=1$, which satisfies the reverse isoperimetric inequality ($R \geq 1$%
) \cite{Isoperimetric}. Notably, a black hole that violates the reverse
isoperimetric inequality, i.e., $R<1$, is called a super-entropic black hole 
\cite{SupperEntropy}.

\section{Conclusions}
In this paper, we combined two nonlinear theories: the
nonlinear massive gravity (dRGT-like massive gravity) as a modified theory
of gravity and the ModMax field as a nonlinear electrodynamics theory. We
obtained exact black hole solutions within this framework and studied the
effects of various parameters of this theory on these black holes, as shown
in Fig. \ref{Fig1}. We found that the number of roots of the metric function
changed with variations in the parameters of the ModMax and massive gravity
theories. These results indicated that the solution could exhibit four
different behaviors: naked singularities, extremal black holes, and
solutions with two or three real roots. Specifically, for large values of
the massive gravity parameters, we encountered multiple horizons when $%
\gamma $ was large. However, for small values of the massive parameter,
increasing $\gamma$ could lead to solutions with two horizons, similar to
ordinary charged black hole solutions.

We obtained thermodynamic quantities such as the Hawking temperature,
electric potential, total mass, total electric charge, and entropy in the
non-extended phase space. These quantities satisfied the first law of
thermodynamics. We studied the effects of parameters from massive and ModMax
theories on these quantities. For example, considering large values of the
massive gravity parameters and varying the ModMax parameter, the number of
roots of the temperature changed. In other words, the temperature could have
one, two, or three roots depending on the value of $\gamma$. In addition.
Furthermore, there were no roots for the temperature when $\gamma$ took on
very large values. Notably, medium black holes in the Maxwell-dRGT-like
massive gravity were non-physical systems due to their negative
temperatures. In contrast, medium black holes in the ModMax-dRGT-like
massive gravity were physical systems because their temperatures were
positive.

The high-energy limit of the total mass of these black holes depended on the
electric charge and the ModMax parameter. The total mass of black holes in
massive gravity was zero for very large values of $\gamma$, a result of the
ModMax theory parameter. Our findings also revealed that the asymptotic
limit of the mass depended on the cosmological constant and the parameters
of dRGT-like massive gravity. In addition, for large values of the massive
gravity parameters, there were three critical points (two minima and one
maximum) for total mass. However, for small values of the massive gravity
parameters, the total mass of these black holes reached a minimum point.

The total electric charge of black holes in ModMax-dRGT-like massive gravity
depended on the ModMax parameter and vanished as $\gamma$ increases.
Additionally, the electric potential and entropy were independent of the
ModMax and massive parameters. Finally, we indicated that the obtained
conserved and thermodynamics quantities satisfied the first law of
thermodynamics.

To investigate the local stability of charged black holes
within the ModMax-dRGT-like massive gravity theory, we plotted the heat
capacity of these black holes in Figs. \ref{Fig4} and \ref{Fig5}. Our
results indicated that large black holes are always stable and considered
physical objects. In contrast, the stability and physicality of medium black
holes depend on the values of the massive and ModMax parameters. For
instance, with large values of the massive gravity parameters and a small $%
\gamma$, both medium and large black holes are stable and physical. However,
at high values of $\gamma$, only large black holes remained stable and
physical. Additionally, when the massive gravity parameters are small and $%
\gamma$ is large, both medium and large black holes are stable and physical.

We expanded our study to calculate various conserved and thermodynamic
quantities of these solutions in the extended phase space. To achieve this,
we used an analogy between thermodynamic pressure and the negative
cosmological constant, allowing us to rewrite all thermodynamic quantities
accordingly. We demonstrated that these quantities satisfied the first law
of thermodynamics, the Smarr relation, and the reverse isoperimetric
inequality (with $R=1$).

The investigation of $P-V$ criticality and phase transitions of black holes in extended phase space \cite{PVI,PVII,PVIII,PVIV,PVV,PVVI,PVVII,PVVIII,PVIX,PVX,PVXI,PVXII,PVXIII} is a fascinating topic that may provide valuable insights into the impacts of ModMax nonlinear electrodynamics and massive gravity on these phenomena. We will explore this further in future work.


\begin{acknowledgements}
We sincerely appreciate the valuable comments from the reviewer, which helped us improve the quality of our work. We would like to thank University of Mazandaran.
\end{acknowledgements}

\end{document}